\documentclass[a4paper,11pt]{article}

\usepackage{mathtools}
\usepackage{amsfonts}
\usepackage{amsthm}
\usepackage{graphicx}

\newcommand{\D}{\mathcal{D}}
\renewcommand{\L}{\mathcal{L}}
\newcommand{\scalar}[2]{\langle#1\,,#2\rangle}
\newcommand{\norm}[1]{\|#1\|}

\renewcommand{\d}{\mathrm{d}}
\newcommand{\pO}{\partial \Omega}
\renewcommand{\H}{\mathcal{H}}

\newtheorem{theorem}{Theorem}

\theoremstyle{definition}
\newtheorem{hypothesis}{Hypothesis}

\newtheorem{definition}[theorem]{Definition}
\newtheorem{example}[theorem]{Example}

\theoremstyle{remark}

\title{BOUNDARY DYNAMICS AND TOPOLOGY CHANGE IN QUANTUM MECHANICS}

\author{\textsc{J.M.~P\'erez-Pardo} \thanks{juanma@na.infn.it }\\ \small   Sezione di Napoli, Istituto Nazionale di Fisica Nucleare, \\ \small
Via Cintia Edificio 6, 80126 Napoli, Italy\\ \and
\textsc{M. Barbero-Li\~n\'an}\thanks{mbarbero@math.uc3m.es}, \textsc{A. Ibort}\thanks{albertoi@math.uc3m.es}\\
\small
Departamento de Matem\'aticas, Universidad Carlos III de Madrid, \\ \small Avenida de la Universidad 30, 28911 Legan\'es, Madrid, Spain
\\ \small and Instituto de Ciencias Matem\'aticas (CSIC-UAM-UC3M-UCM)  \\ \small C/Nicol\'as
Cabrera 13-15, 28049 Madrid, Spain
}

\begin{document}

\maketitle

\begin{abstract}
We show how to use boundary conditions to drive the evolution on a Quantum Mechanical system. We will see how  this problem can be expressed in terms of a time-dependent Schr\"{o}dinger equation. In particular we will need the theory of self-adjoint extensions of differential operators in manifolds with boundary. An introduction of the latter as well as meaningful examples will be given.
It is known that different boundary conditions can be used to describe different topologies of the associated quantum systems. We will use the previous results to study how this topology change can be accomplished in a dynamical way.

\vspace{3mm}

\end{abstract}

\section{Introduction} \label{sec:intro}

A first motivation for the study of quantum systems with boundary is that it is almost necessary to deal with the boundaries of the system when one is describing the interaction of a system with the outer world. Even if the system is considered to be isolated at some region in space, one necessarily needs to deal and describe the interaction of the system with its walls. In addition to this, quantum systems with boundary are getting increasing attention from the scientific community since several interesting physical phenomena are related with the presence of boundaries, like Casimir effect, Quantum Hall effect or Topological insulators \cite{plunien86, morandi88, hasan10, asorey13}.

What will be more important for us in the context of the present article is that boundary conditions are an effective way of describing different topologies. We will devote the rest of this section to give a brief introduction to quantum mechanics which will also serve to fix the notation. In Section \ref{sec:LB} we will introduce the Laplace-Beltrami operator and will show that different boundary conditions do characterise different self-adjoint extensions of this operator. This identification is used in Section \ref{sec:bd} to set the ground for a proper description of dynamical change of the boundary data. Finally, in Section \ref{sec:dtc} we address the problem of changing the topology of a quantum system in a dynamical way.\\

The space of states in quantum mechanics is provided by a complex separable Hilbert space $\H$\,. The scalar product is denoted by $\scalar{\cdot}{\cdot}$ and $\norm{\cdot}$ stands for the corresponding norm.
%
%
Later on we will discuss time-dependent problems. To establish the main notions needed to address them, let us first introduce the time-independent problems. Dynamics in Quantum Mechanics is governed by the Schr\"odinger equation:
$$i\hbar\frac{\d}{\d t}\Psi(t)=H \Psi(t)\,,$$
where $H$ is a linear, self-adjoint, generally unbounded operator acting on the Hilbert space of the system. This operator is called the Hamiltonian operator. If it does not depend explicitly on the time $t$, Schr\"odinger's equation is said to be time-independent. Solutions of the time-independent Schr\"odinger equation can be obtained by means of strongly continuous one-parameter unitary groups.

\begin{definition}
A strongly continuous one-parameter unitary group is a one parameter family of unitary operators $U:\mathbb{R}\to \mathcal{U}(\H)$ such that
$$U(t_1)U(t_2)=U(t_1+t_2)\;,\quad U(0)=\mathbb{I}_\H\;,$$
$$\lim_{t\to 0}\norm{U(t)\Psi-\Psi}=0\;,\quad\forall\Psi\in\H\;.$$
\end{definition}


Stone's theorem establishes a  one-to-one correspondence between strongly continuous one-parameter unitary groups and self-adjoint operators. This correspondence can be summarised by the expression:
$$U(t)=\exp(i t H)\;.$$
Given an initial state $\Psi_0\in\D(\H)$, a solution of Schr\"odinger's equation is $\Psi(t)=U(t)\Psi_0\,.$

%

The operator $H$ is generally an unbounded operator. On the contrary to what happens in the bounded case, there is a difference between symmetric and self-adjoint operators. We will illustrate this difference by means of the following example.

\begin{example}[The momentum operator on the interval {$[0,1]$}]
We will consider the differential operator defined by $$P=i\textstyle{\frac{\d}{\d x}}\;,$$
acting on the Hilbert space of square integrable functions $\L^2([0,1])$\,. Notice that one needs to select a domain for this operator since there are functions $\Psi\in\L^2([0,1])$ that do not verify $P\Psi\in\L^2([0,1])$. For instance, $\Psi=\sqrt{x}\,.$ A suitable choice for the domain is $\D(P)=\mathcal{C}_c^\infty\left((0,1)\right)$, the space of smooth functions with compact support contained in the interior of the interval.

Now we want to obtain the domain of the adjoint operator. In order to determine it we will use Green's formula:
$$\scalar{\Psi}{P\Phi}-\scalar{P\Psi}{\Phi}=i\left[ \bar{\Psi}(1)\Phi(1)- \bar{\Psi}(0)\Phi(0)  \right]\;.$$
It is clear that the right hand side vanishes no matter what the values of $\Psi(1)$ and $\Psi(0)$ are, since $\Phi\in\mathcal{C}_c^\infty\left((0,1)\right)\,.$ Hence, the only requirement for $\Psi$ to be in the domain of the adjoint operator is that $P\Psi\in\L^2([0,1])$\,, i.e., that the first derivative must be also a square integrable function. The space of such functions is called the Sobolev space of order one and is denoted as $\H^1([0,1])\,.$ It is clear now that
$$\D(P^\dagger)=\H^1([0,1])\supset\mathcal{C}_c^\infty\left((0,1)\right)=\D(P)\,.$$
\end{example}

Thus, the momentum operator $P$ with domain $\mathcal{C}_c^\infty\left((0,1)\right)$ is symmetric but not self-adjoint. Being symmetric means that the identity
$$\scalar{\Psi}{P\Phi}-\scalar{P\Psi}{\Phi}=0$$
holds for all $\Phi,\Psi\in\D(P)$\,. To be self-adjoint one needs the extra requirement that $\D(P^\dagger)=\D(P)\,,$ what is not satisfied under the conditions of the example. That $\D(P^\dagger)\supset\D(P)$ is not a particular property of the example but the general situation for symmetric operators. If one wants to select a domain in which the momentum operator is self-adjoint one needs to select a larger domain $\D_{s.a.}\supset\D(P)$ such that $\D(P^\dagger)\supset\D_{s.a.}^\dagger=\D_{s.a.}\supset\D(P)$\,. The general characterisation of those operators admitting such a construction was done by von Neumann, cf. \cite{reed72} and references therein. For differential operators however, one shall take a more direct approach in terms of boundary conditions. This can be proved to be equivalent to the former theory \cite{grubb68}. It is then necessary to select a space that makes the boundary term coming from Green's formula vanish. For instance, one can take the family of domains
$$\D(P_\alpha)=\{\Phi\in\H^1[0,1]\mid \Phi(0)=e^{i\alpha}\Phi(1) \}\,,\quad \alpha\in[0,2\pi)\;.$$
For each value of the parameter $\alpha$ the domain above defines a different self-adjoint extension of the momentum operator.

\section{Self-adjoint extensions of the Laplace-Beltrami operator} \label{sec:LB}

We shall consider the dynamics of a quantum non-relativistic free particle moving on a Riemannian manifold. In particular we will consider that the manifold has boundary, so that different topologies can be described. Let $\Omega$ be a smooth, compact, Riemannian manifold with smooth boundary $\pO$. The dynamics of a quantum free particle constrained to the manifold $\Omega$ are governed by the Laplace-Beltrami operator. If we denote the Riemannian metric by $\eta$, this operator can be defined in terms of local coordinates $\{x^i\}$ as 
$$\Delta_\eta=\sum_{j,k}{\frac{1}{\sqrt{|\eta|}}\frac{\partial}{\partial x^j}}\sqrt{|\eta|}\eta^{jk}{\frac{\partial}{\partial x^k}}\;.$$
The Hilbert space of the system shall be the space of square integrable functions on the manifold $\L^2(\Omega)$\,. As in the case of the momentum operator we can use the boundary data to characterise the different self-adjoint extensions. We use small size greek letters to denote the corresponding restriction to the boundary $\varphi:=\Phi|_{\pO}\;.$ We use dotted small greek letters to denote the restriction to the boundary of the normal derivative pointing outwards $\dot{\varphi}:=\frac{\d\Phi}{\d\mathbf{n}}|_{\pO}$\,. 
Green's formula for the Laplace-Beltrami operator can be written using this notation as
$$\scalar{\Phi}{-\Delta\Psi}-\scalar{-\Delta\Phi}{\Psi}=\scalar{\dot{\varphi}}{\psi}_{\pO}-\scalar{\varphi}{\dot{\psi}}_{\pO}\;,$$
where $\scalar{\cdot}{\cdot}_{\pO}$ denotes the scalar product of the Hilbert space of square integrable functions at the boundary, $\L^2(\pO)$\,.
To determine boundary conditions that lead to self-adjoint extensions of the Laplace-Beltrami operator we need to look for maximal subspaces that make the boundary term vanish. The maximality requirement is needed to ensure that the extension is truly self-adjoint and not just symmetric. One can obtain these boundary conditions by means of the Cayley transform, cf. \cite{asorey05,ibort14}, defined by
$$\varphi_+:=\varphi+i\dot{\varphi}\,,\quad \varphi_-:=\varphi-i\dot{\varphi}\;.$$
In terms of these new variables the boundary term becomes 
$$i\scalar{\varphi_+}{\psi_+}_{\pO}-i\scalar{\varphi_-}{\psi_-}_{\pO}\;.$$
It is clear that boundary conditions of the form 
\begin{equation}\label{eq:asoreyeq}
	\varphi_-=U\varphi_+\Leftrightarrow \varphi-i\dot{\varphi}=U(\varphi+i\dot{\varphi})\;,
\end{equation}
where $U\in\mathcal{U}\left(\L^2(\pO)\right)$ is a unitary operator on the Hilbert space of the boundary, make the boundary term vanish. That these are maximal and that every maximal vanishing subspace of the boundary form can be expressed in this way is a well known result \cite{kochubei75,asorey05,ibort14}. One-to-one correspondence between self-adjoint extensions and unitary operators in the Hilbert space of the boundary can only be achieved in dimension one. Fortunately, imposing extra regularity conditions on the unitary operators, one can still describe a wide class of self-adjoint extensions of the Laplace-Beltrami operator. In particular, those exhibited in Ex.~\ref{ex:particleD}, see \cite{ibort14,ibort14b}.

\begin{example}[Free particle on the interval {$\Omega=[0,1]$}]    \label{ex:interval}
The boundary of the manifold is in this case $\pO=\{0,1\}$ and therefore the unitary group acting on the Hilbert space of the boundary is $\mathcal{U}\left((\L^2(\pO)\right)=\mathcal{U}(2)\,.$ The simplest cases are when $U=-\mathbb{I}_{2\times 2}$ and $U=\mathbb{I}_{2\times 2}$, that correspond respectively with Dirichlet and Neumann boundary conditions. If one considers 
\begin{equation}\label{eq:periodicbc}
	U=\begin{pmatrix} 0 & 1 \\ 1 & 0 \end{pmatrix}\;,
\end{equation}
the boundary condition of Eq.~\eqref{eq:asoreyeq} becomes 
$$\Phi(0)=\Phi(1)\;,\quad\Phi'(0)=\Phi'(1)\;.$$
That is, periodic boundary conditions joining both ends of the interval.
On the other hand if one takes 
\begin{equation}\label{eq:quasiperiodicbc}
	U=\begin{pmatrix} 0 & e^{i\alpha} \\ e^{-i\alpha} & 0 \end{pmatrix}\;,
\end{equation}
the boundary condition of Eq.~\eqref{eq:asoreyeq} becomes 
$$\Phi(0)=e^{i\alpha}\Phi(1)\,,\quad \Phi'(0)=e^{i\alpha}\Phi'(1)\;.$$
We call the latter quasi-periodic boundary conditions.
\end{example}

\begin{example}[Free particle in two intervals, $\Omega=I_1\cup I_2$]
In this case the boundary is the set $\pO=\{a_1,b_2,a_2,b_2\}$ (see Fig.~\ref{fig:circles})\,, and the unitary group at the boundary is therefore $\mathcal{U}\left(\L^2(\pO)\right)\simeq\mathcal{U}(4)\,.$ We can now reproduce block-wise the conditions given by the periodic conditions, see Eq.~\eqref{eq:periodicbc}, identifying the endpoints $a_1, b_1, a_2, b_2$ in different ways. For instance, we can consider the following unitaries \cite{bal95}:
	$$U_1=\begin{pmatrix} 0 & 1 & 0 & 0 \\ 1 & 0 & 0 & 0 \\ 0 & 0 & 0 & 1 \\ 0 & 0 & 1 & 0 \end{pmatrix}\;,\quad U_2=\begin{pmatrix} 0 & 0 & 0 & 1 \\ 0 & 0 & 1 & 0 \\ 0 & 1 & 0 & 0 \\ 1 & 0 & 0 & 0 \end{pmatrix}\;.$$ The former corresponds to the topology of two circles whereas the latter corresponds to the topology of one circle.
\begin{figure}[h]
\center
\includegraphics[height=2cm,width=4.5cm]{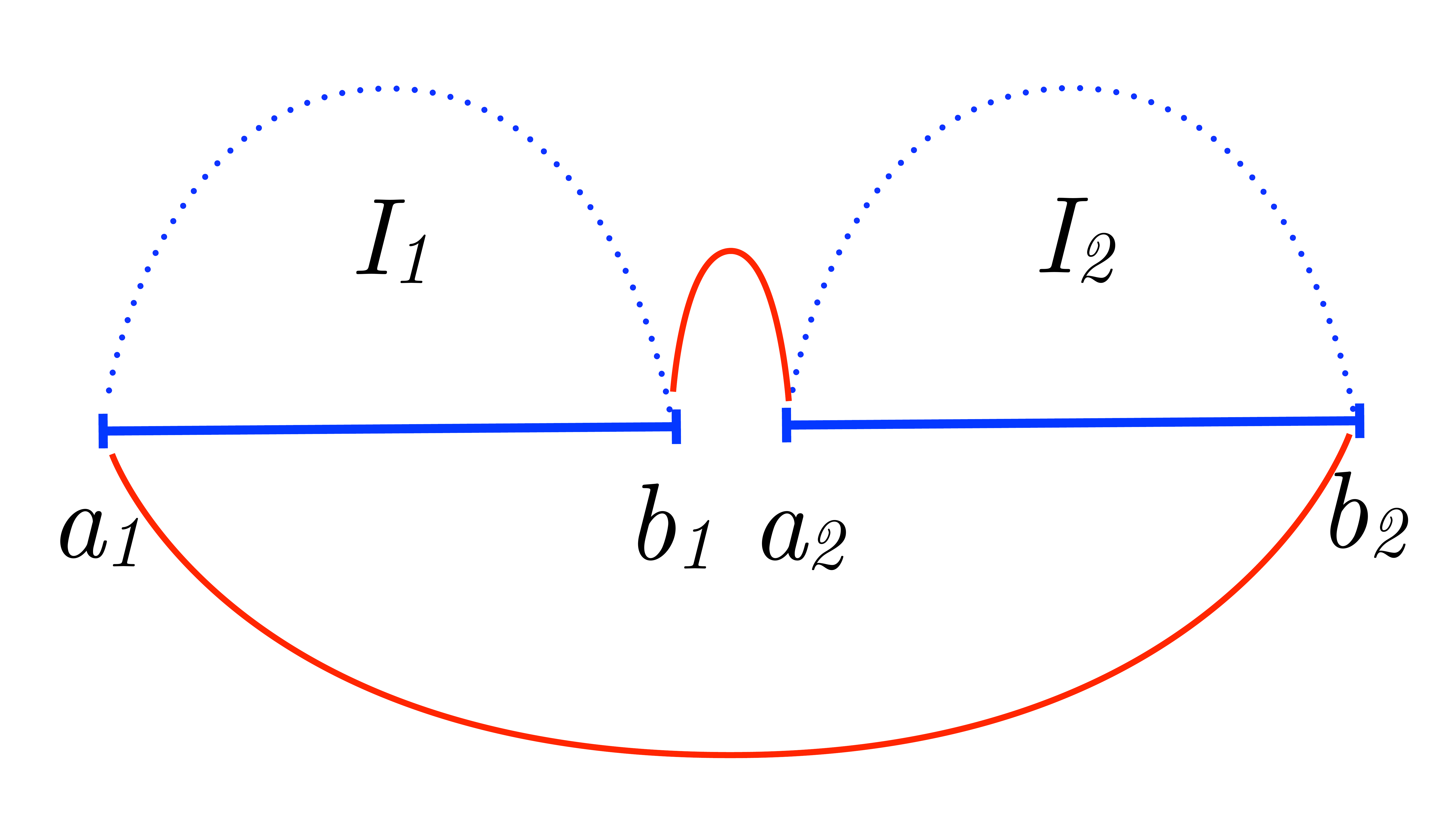}
\caption{Identifications of the endpoints corresponding to different choices of the boundary conditions. Dotted lines correspond to the topology of two circles, $U_1$. Solid lines correspond to the topology of one circle $U_2$\,.}
\label{fig:circles}
\end{figure}
\end{example}

\begin{example}[Free particle on the disk, $\Omega=D$] \label{ex:particleD}
In this case the boundary is given by the unit circle $\pO=S^1$\,. The unitary group at the boundary is therefore given by the infinite dimensional unitary group $\mathcal{U}\left(\L^2(S^1)\right)$\,. To impose different topological structures we shall introduce a splitting of the boundary $\pO=\Gamma_1\cup\Gamma_2$ where $\Gamma_1$ and $\Gamma_2$ are disjoint and diffeomorphic to each other (see Fig.~\ref{fig:particleD}). The isomorphism $\L^2(\pO)\simeq\L^2(\Gamma_1)\oplus\L^2(\Gamma_2)$ allows us to define unitary operators adapted to this block-wise structure. For instance one could select $$U=\begin{pmatrix} -\mathbb{I}_{\L^2(\Gamma_1)} & 0 \\ 0 & \mathbb{I}_{\L^2(\Gamma_2)} \end{pmatrix}\;,$$ what would impose Dirichlet boundary conditions on $\Gamma_1$ and Neumann boundary conditions on $\Gamma_2$\,.

\begin{figure}[h]
\center
\includegraphics[height=2cm]{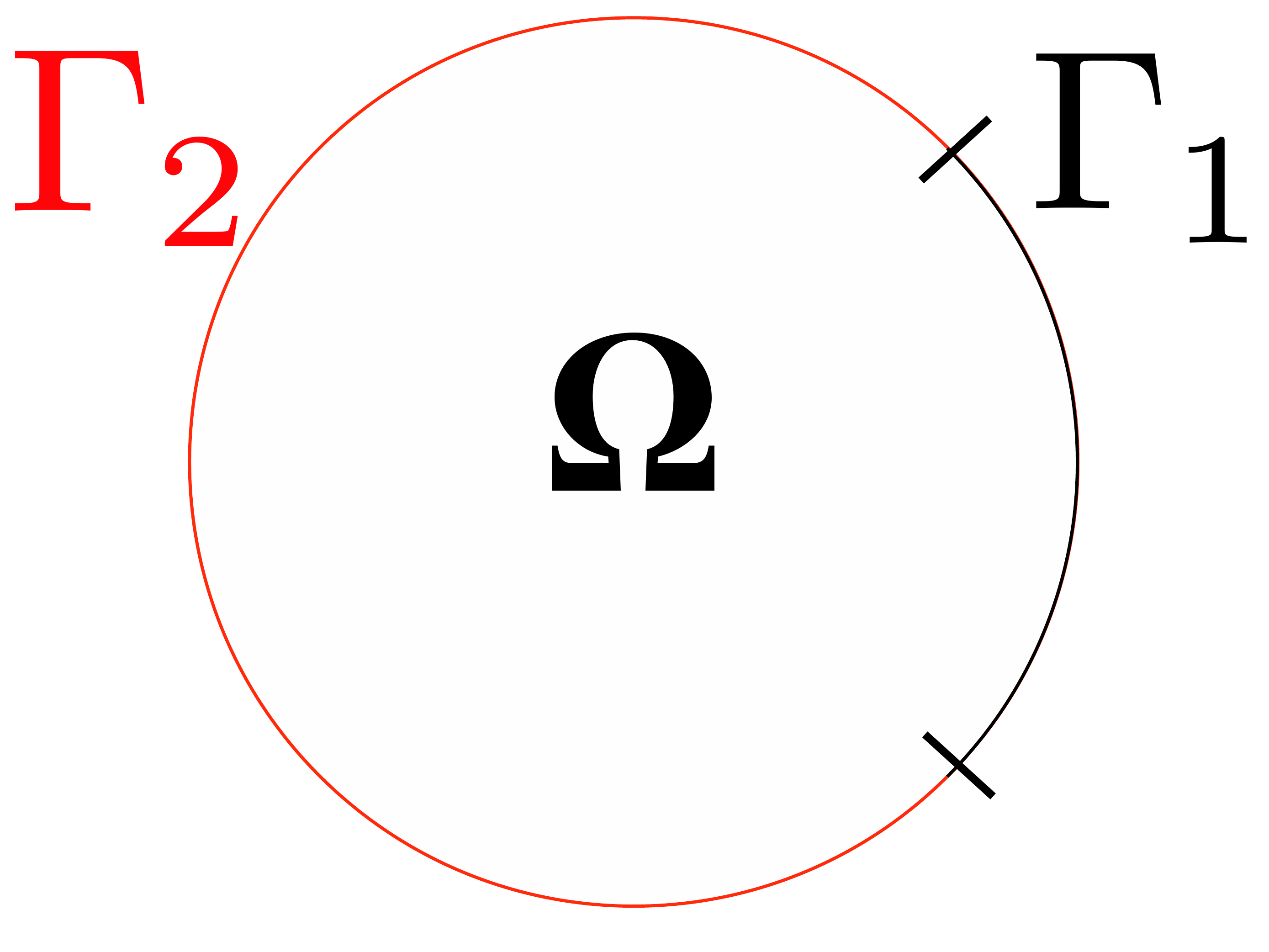}
\caption{The boundary of the manifold, $\pO=S^1$, is split into two parts, $\Gamma_1$ and $\Gamma_2$\,.}\label{fig:particleD}
\end{figure}

\noindent We want to be able to describe different topologies in this situation. From Fig.~\ref{fig:particleD} it is clear that one cannot directly use the structure given in $\eqref{eq:periodicbc}$ since $\Gamma_1$ and $\Gamma_2$ are diffeomorphic as differentiable manifolds but not necessarily isometric. To overcome this difficulty we can do the following. Select a reference manifold $\Gamma_0$, diffeomorphic to $\Gamma_1$ and $\Gamma_2$ but otherwise arbitrary. We denote the diffeomorphisms by $g_i:\Gamma_i\to\Gamma_0$\,. Fix a volume element, $\d\mu_0$, for the manifold $\Gamma_0$\,. Now we can define the following unitary operators $T_i=\L^2(\Gamma_i)\to\L^2(\Gamma_0,\d\mu_0)$
\begin{equation}\label{eq:isomorphisms}
	T_i\Phi:=\sqrt{|J_i|\mu_i}(\Phi\circ g_i)\;,
\end{equation}
where $|J_i|$ is the Jacobian determinant of the diffeomorphism $g_i$ and $\mu_i$ the proportionality factor: $g^*_i\d\mu_\eta=\mu_i\d\mu_0$\,, cf. \cite{ibort14}. Using these unitary operators we can define the following unitary operator on $\L^2(\Omega)$:
\begin{equation}\label{eq:unitaryblock}
U=\begin{pmatrix} 0 & T_1^\dagger T_2 \\ T_2^\dagger T_1 & 0 \end{pmatrix}\;.
\end{equation}
This operator leads to the boundary conditions 
$$T_1(\varphi|_{\Gamma_1})=T_2(\varphi|_{\Gamma_2})\;,\quad T_1(\dot{\varphi}|_{\Gamma_1})=-T_2(\dot{\varphi}|_{\Gamma_2})\;,$$ which can be called generalised periodic boundary conditions.
\end{example}

\section{Boundary Dynamics} \label{sec:bd}

As stated at the beginning of this article we will discuss dynamics of a time-dependent quantum mechanical system, i.e. the Hamiltonian operator will depend explicitly on time. This means that we will have a family of self-adjoint operators $H(t)$\,. In a conventional time-dependent problem,  one typically has a time-dependent perturbation modifying the Hamiltonian corresponding to the free particle. Even more generally, it is usually assumed that the functional form of the Hamiltonian operator depends explicitly on time. In contrast, we aim to modify the boundary conditions externally and leave the functional form of the Hamiltonian operator unchanged. As we have seen in the previous section, this amounts to change the self-adjoint extension of the operator. Hence the Hamiltonian operator will have a fixed functional form but its domain will depend on time through the boundary conditions. More concretely, we will have a set of parameters, $u\in\mathcal{M}$, characterising different boundary conditions that will depend explicitly on time. The space of parameters $\mathcal{M}$ will be in general a subset of the set of all possible self-adjoint extensions of the Hamiltonian operator.

It is time now to discuss the time-dependent Schr\"odinger equation
$$i\hbar\frac{\d}{\d t}\Psi(t)=H(t) \Psi(t)\;.$$
Recall the discussion of Section \ref{sec:intro} where the solutions of the time-independent Schr\"odinger equation where obtained in terms of a strongly continuous one-parameter group of unitary operators. Similarly, solutions of the time-dependent equation can be obtained in terms of what is known as a unitary propagator.
\begin{definition}
A unitary propagator is a strongly continuous two parameter family of unitary operators $U:\mathbb{R}\times\mathbb{R}\to \mathcal{U}(\H)$ that satisfy the properties
$$U(t,s)U(s,r)=U(t,r)\;,\quad U(t,t)=\mathbb{I}_{\H}\;.$$
\end{definition}
Unlike the time-independent case, there is no general result like Stone's theorem guaranteeing the existence of a unitary propagator for a given time-dependent Hamiltonian. Hence, given a family of self-adjoint operators $\{H(t), t\in I\subset \mathbb{R}\}$ one needs to look for sufficient conditions for it to exist.
The situation is particularly involved since the domains of the operators are going to change with time. We will take now some simplifying hypothesis.

\begin{hypothesis}\label{H:1}
Let $\D(u)$ denote the domain of the self-adjoint extension of the operator $H$ defined by the boundary condition determined by $u\in\mathcal{M}$. For every $u\in\mathcal{M}$, the spectrum of $\left( H, \D(u) \right)$ only contains eigenvalues with finite degeneracy. That means that we do not consider situations with singular nor continuous spectrum. This implies that for each $u\in\mathcal{M}$ there exists a complete orthonormal set of eigenvalues $$\{\Phi_u^n\}_{n\in\mathbb{N}}\;.$$
\end{hypothesis}
Under these conditions a reference self-adjoint extension $\left( H,\D(u_0) \right)$ with orthonormal base $\{\Phi_0^n\}_{n\in\mathbb{N}}$ can be fixed. We define a family of unitary operators $V_u:\H\to\H$ by 
$$V_u:\Phi_u^n\to\Phi_0^n\;.$$
This definition characterises completely the unitary operators $V_u$\,. We will also require the following.

\begin{hypothesis}\label{H:2}
For each $u\in\mathcal{M}$ the unitary operator $V_u$ must be a bijection $$V_u:\D(u)\to\D(u_0)\;.$$
\end{hypothesis}

Notice that this is not an immediate consequence of the definition of $V_u$\,. In fact, this hypothesis imposes asymptotic conditions on the sequences of eigenvalues of the family $\left(H,\D(u)\right)$\,.

Under conditions \ref{H:1} and \ref{H:2} the operator $V_uHV_u^\dagger$ is a self-adjoint operator with domain $\D(U_0)$\,. Let $\Psi^0:=V_u\Psi\in\D(u_0)$ for any $\Psi\in\D(u)\,.$ Now we can compute Schr\"odinger's equation in terms of $\Psi^0$ to get:
\begin{equation}\label{eq:fixedD}
i\hbar \frac{\d}{\d t}\Psi^0= V_uH V_u^\dagger\Psi^0 - i\hbar V_u \dot{V}_u^\dagger\Psi^0\;.
\end{equation}
To obtain this equation we have used Leibniz rule. The derivatives have to be understood in the sense of the strong topology. In the right hand side $\dot{V}_u^\dagger$ stands for the derivative of the family of unitary operators where the parameters $u=u(t)$ are assumed to be time-dependent. This derivative will be in general an unbounded operator and might even not exist depending on the particular problem. In order to guarantee the existence of a unitary propagator that solves the Schr\"odinger equation we can use the following particular form of a theorem due to J.~Kisynski, cf.\ \cite{kisynski63}.

\begin{theorem}
Under the hypotheses \ref{H:1}, \ref{H:2} and:
\begin{hypothesis}\label{H:3}
$\scalar{\Psi}{V_u^\dagger\Phi}$ is twice continuously differentiable for all $\Phi,\Psi\in\H$\,.
\end{hypothesis}
\begin{hypothesis}\label{H:4}
$\scalar{\Psi}{V_u H V_u^\dagger\Phi}$ is continuously differentiable for all $\Psi\in\H$ and for all $\Psi\in\D(u_0)$\,.
\end{hypothesis}
\begin{hypothesis}\label{H:5}
$\left(H, \D(u(t))\right)$ is semibounded from below uniformly.
\end{hypothesis}
There exists a unique unitary propagator solving the time-dependent Schr\"{o}\-dinger equation defined by the time-dependent Hamiltonian $H$.
\end{theorem}

From the form of Eq.~\eqref{eq:fixedD} one can see why the regularity and differentiability conditions are needed. The semibounded condition is a further spectral condition on the family of self-adjoint operators. In fact, the condition considered originally by Kisynski is more general and includes this one, that turns out to be enough in our case. Under certain conditions on the family of unitary operators $U$ describing the boundary condition and therefore parametrising the different self-adjoint extensions, the Laplace-Beltrami operator can be proved to be semibounded. We refer to \cite{ibort14} for a detailed analysis of this situation.

\begin{definition}
	We say that a unitary operator $U\in\mathcal{U}\left(\L^2(\pO)\right)$ has spectral gap if one of the following conditions hold.
	\begin{itemize}
		\item $U+\mathbb{I}$ is invertible
		\item $\{-1\}$ is an element of the spectrum of $U$ and is not an accumulation point.
	\end{itemize}
\end{definition}

\begin{theorem}[Ibort, Lled\'{o}, P\'{e}rez-Pardo, \cite{ibort14}]
Let U be a unitary operator with spectral gap. Then the self-adjoint extension of the Laplace-Beltrami operator defined by U is semibounded from below.
\end{theorem}

The lower bound is proved to be directly related to the gap of the unitary operator at the boundary, that is, the distance between $\{-1\}$ and the closest element of the spectrum to $\{-1\}$\,. In order to have a uniform semibound for the family of self-adjoint extensions it is enough to have a uniform gap for the family of unitary operators.

\begin{example}[Laplacian with quasi-periodic boundary conditions]\label{ex:faraday}
Let us consider the Laplace-Beltrami operator $H=-\frac{\d^2}{\d\theta^2}$ on the interval $\Omega=[0,2\pi]$\,. We consider the one-parameter family of self-adjoint extensions given by quasi-periodic boundary conditions defined in Ex.~\ref{ex:interval} by the unitary operator \eqref{eq:quasiperiodicbc}. This family of operators satisfies condition \ref{H:5} since all the elements have spectral gap. This is so because the spectrum of every element of the family is $\sigma(U)=\{-1,1\}$\,. The eigenvalues and eigenvectors can be computed explicitly and one gets 
$$E^n_\epsilon=(n+\epsilon)^2\;,\quad \Phi_n^\epsilon(\theta)=\frac{1}{\sqrt{2\pi}}e^{i(n+\epsilon)\theta}\;,n\in \mathbb{Z}\;.$$
The family of unitary operators becomes in this case $$V_{\epsilon}\Psi:=e^{-i\epsilon\theta}\Psi\;.$$ That is, a family of unitary multiplication operators acting on $\L^2([0,2\pi])$\,. It is clear that \ref{H:1} and \ref{H:2} are also satisfied. Further calculations show that $$\dot{V}^\dagger_\epsilon=i\theta\dot{\epsilon}e^{i\epsilon\theta}\;.$$ Hence, the regularity conditions \ref{H:3} and \ref{H:4} hold if the first and second derivatives of the function $\epsilon:\mathbb{R}\to[0,2\pi]$ are bounded. Substituting all the expressions into the time-dependent Schr\"odinger equation \eqref{eq:fixedD} it is straightforward to get
$$i\hbar\frac{\d}{\d t}\Psi=\left[i\textstyle{\frac{\d}{\d\theta}}-\epsilon \right]^2\Psi+\hbar\theta\dot{\epsilon}\Psi\;,\quad\Psi\in\D(\epsilon=0)\;.$$
This equation corresponds physically to the quantum Faraday law. It represents the motion of a particle spinning in a planar wire surrounding a magnetic flux of intensity $\epsilon$ that changes with time. The derivative of this magnetic flux is proportional to the electric potential that the particle experiences.
\end{example}

\section{Dynamical topology change} \label{sec:dtc}

\begin{figure}[h]
\center
\includegraphics[height=2cm]{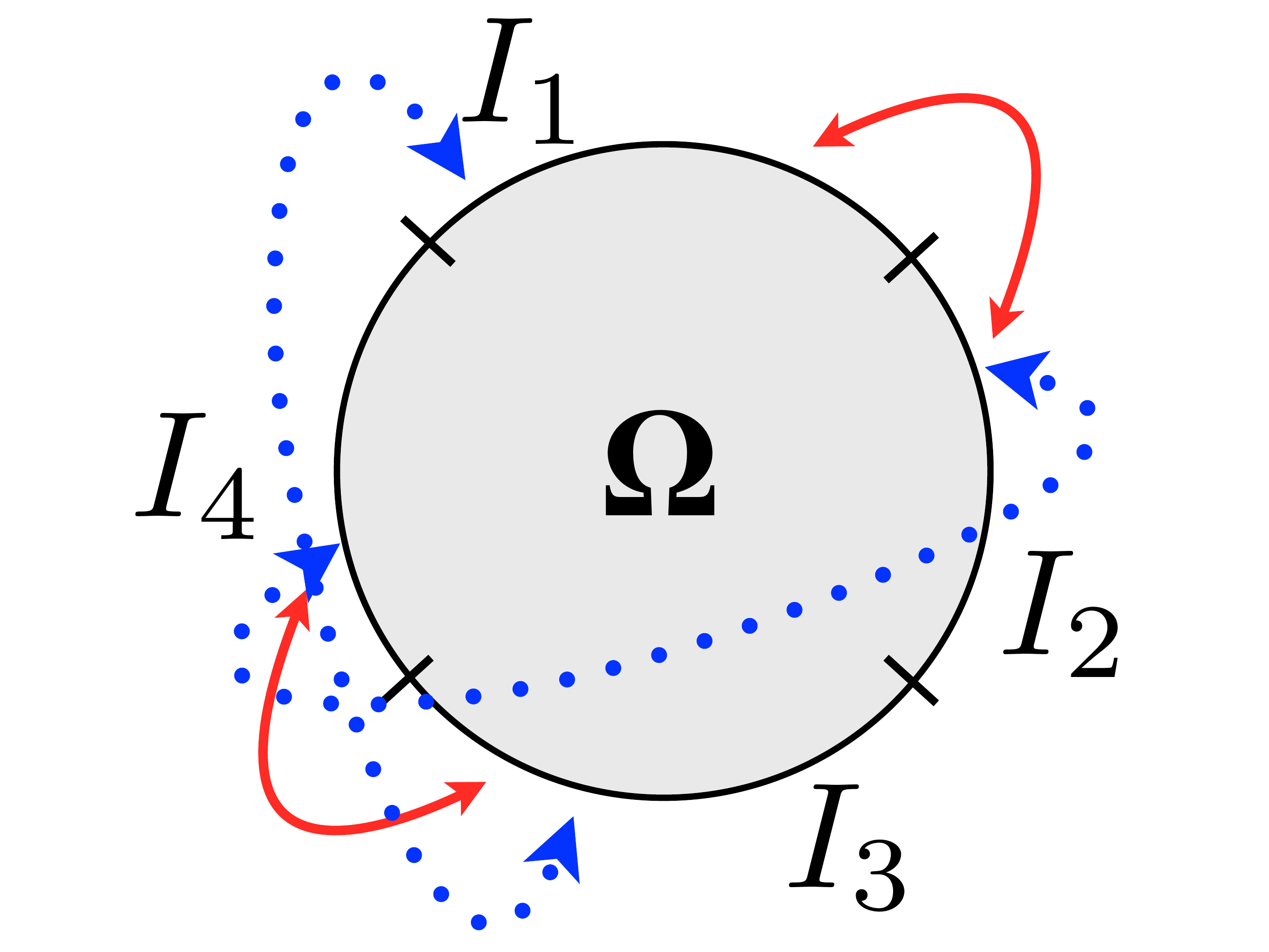}
\caption{Disk with boundary conditions split into four diffeomorphic regions. Identification along the solid lines correspond to the topology of a sphere. Identification along the dotted lines corresponds to the topology of a torus.}\label{fig:disk4i}
\end{figure}

In Section~\ref{sec:LB} we have shown how we can use boundary conditions to define self-adjoint extensions of the Laplace-Beltrami operator, whose domains present different topological structures. In Section~\ref{sec:bd} we have addressed the problem of existence of solutions of the time-dependent Schr\"odinger equation when the time dependence is located strictly at the boundary conditions. We also provide a set of sufficient conditions (\ref{H:1} to \ref{H:5}) for solutions to exist. Now we will study if we are able to paste everything together to achieve a dynamical variation of the topology. Let us work a further example and check whether or not these conditions are met. We shall consider the same construction introduced in Ex.~\ref{ex:particleD}. This time we split the boundary into four different pieces,  $I_1$, $I_2$, $I_3$ and $I_4$ (see Fig.~\ref{fig:disk4i}).
We can use the isomorphism \eqref{eq:isomorphisms} constructed in Ex.~\ref{ex:particleD} to identify the pieces in different ways. The unitary operators defining the boundary conditions will have block-wise structure with blocks given by \eqref{eq:unitaryblock}. We consider two limiting cases for the time-dependent dynamics:
\begin{itemize}
\item Topology of a sphere.
\item Topology of a torus.
\end{itemize}
We need to select a path in the space of self-adjoint extensions that connects these two cases. Regarding Hyp.~\ref{H:5} it is easy to check, cf.\ \cite[Corollary 3.5]{ibort13}, that when the spectrum of the unitary operator is of the form $\sigma(U)=\{-1,1\}$\ the spectrum of the Laplace-Beltrami operator is non-negative. Moreover, in this situation one can use spectral bracketing techniques, cf. \cite{lledo08}, to show that the $n-th$ eigenvalue of any extension parametrised by an unitary operator with spectrum $\sigma(U)=\{-1,1\}$ satisfies
$$\lambda_n^N\leq\lambda_n(u)\leq \lambda_n^D\;,$$
where $\lambda_n^N$ and $\lambda_n^D$ are the $n$-th eigenvalues of the Laplace-Beltrami operator with Neumann boundary conditions and Dirichlet boundary conditions respectively. This shows that \ref{H:1} and \ref{H:2} hold. Now we need to address the regularity conditions. A sufficient condition for \ref{H:3} and \ref{H:4} to hold would be that the eigenvalues and eigenfunctions were smooth functions with respect to $t$ and that the norm of the first and second derivatives could be bounded by a constant that does not depend on $n$. This is exactly the situation in Ex.~\ref{ex:faraday}. Whether or not this condition is achieved will depend on the path chosen. The existence of such a path is part of actual research and we leave the analysis for future work.

\section*{Acknowledgments}
JMPP wants to thank the organisers and the committee of the XXIII International Fall Workshop on Geometry and Physics for their kind invitation and support. He was partly supported by the project MTM2010-21186-C02- 02 of the spanish Ministerio de Ciencia e Innovaci\'{o}n.

\end{document}